# Lifetime Limited and Tunable Quantum Light Emission in h-BN via Electric Field Modulation

Authors: Hamidreza Akbari[1], Souvik Biswas[1], Pankaj K. Jha[1], Joeson Wong[1], Benjamin Vest[1,2], Harry A. Atwater[1]

[1]Thomas J. Watson laboratory of applied physics, California Institute of Technology, Pasadena, CA., 91106

[2] Université Paris-Saclay, Institut d'Optique Graduate School, CNRS, Laboratoire Charles Fabry, 91127, Palaiseau, France.

Abstract:

Color centered-based single photon emitters in hexagonal boron nitride (h-BN) have shown promising photophysical properties as sources for quantum light emission. Despite significant advances towards such a goal, achieving lifetime-limited quantum light emission in h-BN has proven to be challenging, primarily due to various broadening mechanisms including spectral diffusion. Here, we propose and experimentally demonstrate suppression of spectral diffusion by applying an electrostatic field. We observe both Stark shift tuning of the resonant emission wavelength, and emission linewidth reduction nearly to the homogeneously broadened lifetime limit. Lastly, we find a cubic dependence of the linewidth with respect to temperature at the homogeneous broadening regime. Our results suggest that field tuning in electrostatically gated heterostructures is promising as an approach to control the emission characteristics of h-BN color centers, removing spectral diffusion and providing the energy tunability necessary for integrate of quantum light emission in nanophotonic architectures.

Hexagonal boron nitride (h-BN) is a van der Waals material with a large band gap of ~6 eV. It has played a pivotal role in the development of 2D materials-based devices as an excellent gate dielectric and atomically smooth substrate [1,2]. This material also hosts single photon emitters in the visible range of the electromagnetic spectrum [3,4]. Color centers in h-BN exhibit high stability at room temperatures and above [5,6], as well as high Debye-Waller factors [6–8], and potentially complete mechanical decoupling of the emitter from vibrations of the lattice [9,10]. The atomistic picture of the origin of these emitters is still under debate, [11–14] but the role of carbon in the formation of emitters at visible range of spectrum has been established [15]. Optically driven magnetic resonance of ensemble [16,17] and single [18] emitters in h-BN shows these spins can be magnetically addressed, useful for sensing and quantum computation applications.

Control of many quantum optical processes requires a source of spectrally indistinguishable photons [19,20]. Thus narrow, ideally lifetime limited emission linewidth is therefore a crucial issue for such applications. While color center point defects in h-BN are robust and bright sources of single photons even at room temperature, the emission spectrum is typically strongly affected by various broadening mechanisms. Previous studies have identified spectral diffusion as one of the major broadening for h-BN single photon sources, even at cryogenic temperatures [21–24]. Spectral diffusion corresponds to rapid fluctuations in the energy of the emitter which manifests itself as rapid movement of narrow homogeneously broadened spectral lines. As a result, the observed line profile is a convolution of the homogenously broadened spectral line and inhomogeneous spectral diffusion driven broadening. ($\gamma = \gamma_{hom} + \gamma_{inhom}$) The homogenously broadened linewidth depends on the lifetime and dephasing processes ($\gamma_{hom} = \frac{1}{2\pi T_1} + \frac{1}{\pi T_2^*}$). Here, $T_1$ is the decay lifetime of the emitter and $T_2^*$ is the pure dephasing time scale which can result from spin-bath and/or emitter-phonon interactions. When the single photon emission is purely lifetime limited, the inhomogeneous broadening is eliminated and the pure dephasing contribution to the homogeneous linewidth is zero, so the lifetime limit of the linewidth is $\frac{1}{2\pi T1}$.

While the microscopic origin of spectral diffusion broadening is still under debate, there is growing consensus that itinerant localized charges are a source of time-dependent electric fields in the emitter environment, which modulates the dipole overlap wavefunction leading to spectral diffusion. Photoionization of nearby impurities [23,25], charging and discharging of nearby charge traps [23,25], and presence of mobile charged impurities [22] are some potential sources of these localized charges. The time scale of spectral diffusion can range from microseconds [22] to milliseconds [23,24] to even seconds and minutes, depending on the dominant source of fluctuating environmental and stray fields [26]. Also, the dependence of spectral diffusion on pump power [24] supports the idea that the charges responsible for spectral diffusion can be photo-activated. So far, several studies have attempted different methods to suppress spectral diffusion in h-BN: for example, one study used anti-Stokes excitation to stabilize the spectral line [26], but linewidth narrowing was not studied. Another study used transparent conducting substrates and reduced the inhomogeneous linewidth by 45% [21] but the resulting linewidth narrowing is still far from the lifetime limit. Another study showed that by combining a resonant excitation and lowering the excitation power, the effect of spectral diffusion was reduced [24]. In quantum dot single emitters, use of electric field for charge depletion is established as a method to increase indistinguishability of the emitter [27].

In this report, we present a new method to suppress spectral diffusion of an h-BN emitter and narrow the linewidth by nearly two orders of magnitude, approaching the lifetime limit of the emitter linewidth. To achieve this goal, we apply an out of plane electrostatic field to an emitter in h-BN and study its absorption spectra via photoluminescence excitation (PLE) spectroscopy.

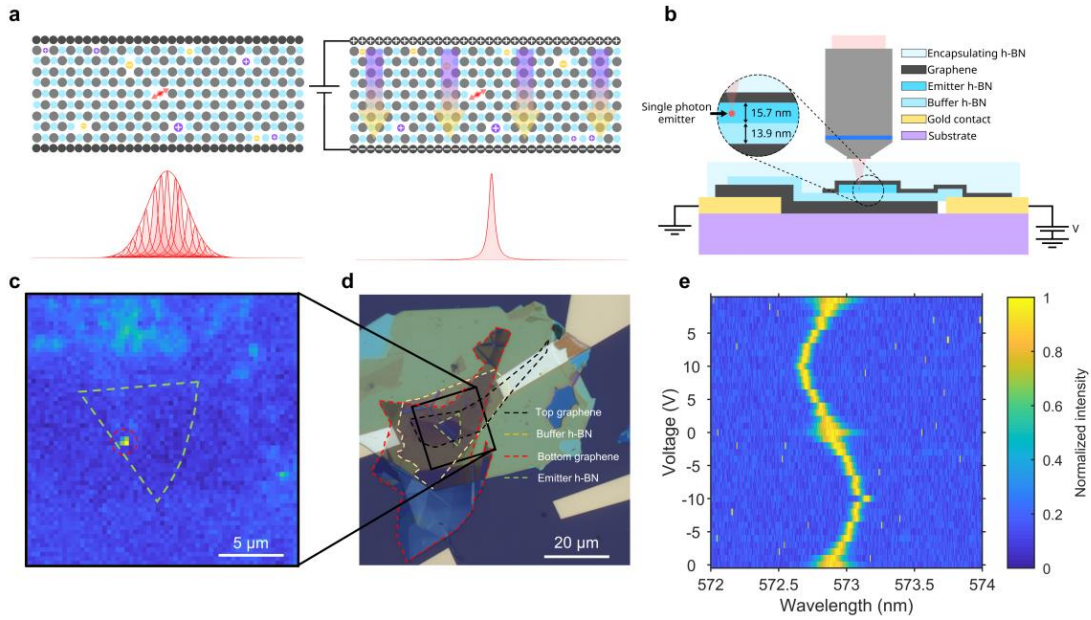

**Figure 1**. (a)Schematic of the effect of electrostatic field on the single photon emitter, top panels show the emitter located in the h-BN crystal with and without the electric field and the bottom panels show the emission spectra. Electric field locks charges located near the surface of h-BN at their place and reduces spectral diffusion. (b) Schematic of the device consisting of top and bottom graphene contacts, h-BN contataining emitter (shown as the red dot) and buffer h-BN.(c)PL map of the region containing the emitter. Each pixel is the integrated spectra from 570nm to 575nm (d). Microscope image of the device flake edges are marked with dashed lines show location of top and bottom graphene and buffer and emitter h-BN. (e) Normalized photoluminescence spectra of emitter ZPL at T=6.5K at voltages ranging form -10V to 10 V. D. PL map of the device, the white circle shows emitter's position.

For this experiment, an h-BN emitter with a zero-phonon line (ZPL) at 573 nm is studied. The emitter is induced in h-BN by the carbon doping method explained in the methods section; a Voigt fit to the line shape reveals a ZPL linewidth of 0.90 meV at 6.5 K, which is higher than the average linewidth at cryogenic temperatures for emitters in thick exfoliated h-BN flakes [21], suggesting a significant broadening due to spectral diffusion.

The emitter is placed inside an all van der Waals device consisting of few-layer graphene electrodes and an additional buffer h-BN layer to prevent electrical shorting between the two electrodes as shown in Fig. 1-b. Second-order intensity autocorrelation experiment shows a $g^{(2)}(0)=0.14$ indicating single photon emission, while time resolved photoluminescence measurement reveals a decay lifetime of 3.36ns which corresponds to 47 MHz lifetime limited linewidth as shown in Fig. 2. The sample is cooled down to 6.5K and the photoluminescence (PL) of the emitter is studied at voltages from -10V to 10V at steps of 1 V, under 532 nm excitation. As can be seen in Fig. 1-e, the Stark effect produces a linear change in the energy of the ZPL, which suggests a non-zero out of plane dipole moment. The slope of peak position as a function of voltage in both PL (Fig. 1-e and Fig. S-3) and PLE (Fig. 3) measurements correspond to a value of 2.7 meV/(V/nm), considering the 29.6 nm thickness of h-BN.

In other studies, a Stark tunability of 43 meV/(V/nm) [28] for in-plane electric field, and 2.5-15 meV/(V/nm) [29] for out of plane electric field is observed for h-BN emitters. Comparing with our result and assuming the dipole moment of different h-BN emitters to be comparable, we conclude that our emitter should be mainly in-plane with a small out of plane component. Polarization measurement of the emitter (Fig. S-2) also shows a linearly polarized emission in accord with a small out of plane dipole.

More notably we see the linewidth narrowing as the voltage increases even in the non-resonant excitation regime (PL measurements). As the absolute value of voltage increases the linewidth approaches the limits of spectrometer resolution equal to 0.2 meV. We note that as the emitter undergoes several cycles of voltage tuning the broadening at 0V remains unchanged suggesting negligible hysteresis in this process. This lack of hysteresis reveals that the charges responsible for spectral diffusion are not permanently removed/screened by the electric field. This supports the idea that generation and recombination of the charges responsible for spectral diffusion is a dynamic process, without long-term charge trapping, and that this dynamic process can be negated at high enough electrostatic field. This effect might be caused by the excitation laser, charging the surrounding dark defects, or mobilizing the charges in the surrounding of the emitter.

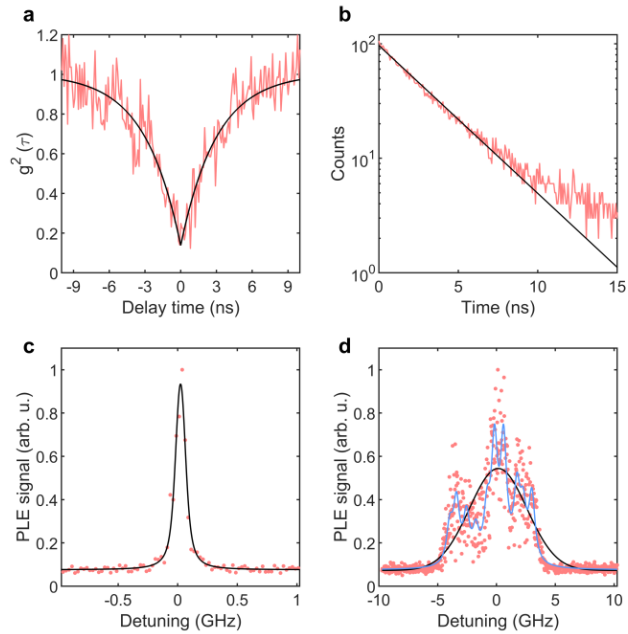

**Figure 2**. (a) Second order autocorrelation function for photon emission. The fit has a value of 0.14 at τ=0. (b) Time resolved photoluminescence of the emitter. The solid line corresponds to single exponential fit with time constant of 3.36ns this value corresponds to lifetime limited broadening of 47 MHz. The deviation of data from fit at t>10ns is a result of hitting the noise floor of the photodetector. (c) Normalized cryogenic (T=6.5K) PLE spectrum of the emitter at V=10V shows linewidth of 89 MHz. (d) At V=0V PLE spectra shows multiple peaks. The blue trace shows a fit with 10 Voigt functions with the same line shape as part (c), the black trace shows a single Voigt fit to assign a FWHM value equal to 8.5 GHz.

To further study the effect of electrostatic fields, without the limitations imposed by the resolution of grating-based spectroscopy and non-resonant excitation, we performed photoluminescence excitation spectroscopy (PLE) experiments using a continuous wave, tunable, narrow linewidth laser source. In PLE experiments, a tunable source of CW laser excitation is slowly scanned across the zero-phonon line and the intensity of the phonon sideband is measured as a function of laser detuning. As a result, this method can measure the absorption spectra of ZPL with a resolution approaching the linewidth of the scanning laser source (~100 kHz). As shown in Fig. 3, a two orders of magnitude reduction in linewidth is observed with the application of an electrostatic field, almost completely removing the inhomogeneous broadening effects at V=10V with a fitted linewidth of 89 MHz as shown in Fig 2-C. The difference between the lifetime limit of 47 MHz and the measured value for linewidth can have several possible explanations. One possible origin of this difference can be residual spectral diffusion, while the electric field has suppressed most of the effect of local charges that impose spectral diffusion, it is possible that some local charge dynamic still exists and the energy barrier to suppress these fluctuations are larger than the energy provided by the electric field. A more likely source of broadening can be dephasing processes. In such a case the observed linewidth will be limited by the total decoherence time scale consisting of lifetime and dephasing time scales. The source of this dephasing can be lattice vibrations, in which case a further reduction of the temperature can help to close the gap and achieve lifetime limited emission.

The line shape of emitter at V=0V (Fig 2-d) and voltages near zero (Fig 3-a) exhibits several peaks. This behavior is expected from spectral diffusion broadening as it is caused by the spectral line jumping between several different energies. As PLE itself has a time scale comparable to the time for only few jumps, the PLE spectrum does not show a single wide peak. The result of fitting with sum of 10 Voigt functions to V=0V data can be seen in Fig 2-d. This behavior makes it challenging to assign a value for FWHM. So, we decided to assign a FWHM to these peaks by simply fitting a single Voigt function to the data and report the FWHM of the fitted peak. Even though this method does not capture the full complexity of the spectrum, it is effective in distinguishing between narrow and wide spectra and assigning a value for FWHM. The result is shown in Fig 3-b.

It is worth noting that in recent studies, [30,31] of the effect of electrostatic field on h-BN emitters, it has been observed that the emitter charge state can be modulated because of electrostatic doping, such that the emitter can switch between optically dark and bright states. However, we do not observe such an effect here, which can be understood as follows. These studies employed graphene electrodes in contact with the emitter-hosting h-BN flake. We suggest that for emitters close to the hBN surface, application of voltage induces electron tunneling through the h-BN barrier, thus populating, and depopulating the emitter states in the near-surface region (within a few monolayers), effectively modulating the optical transitions from off to on. In the present work, a second h-BN layer was added to prevent electrical shorting between the two electrodes, and but another consequence of this layer configuration is that it prevented electron tunneling, since the emitter is not in close contact with the electrodes. Our results suggest that the emitter is not within a tunnelling distance from the bottom electrode, implying also that it is either the middle of closer to the top of the flake. The spatial distribution of h-BN emitters has not been very well studied, although several reports asserted that emitters are mostly located in the vicinity of edges of the flake. [32,33] Also, one study on axial location of emitters in h-BN has shown them to be close to the surface of a 300nm thick flake of h-BN [34]. So, it is possible that in most cases the emitters are either located near the surface or close to edges of the flake, which supports our understanding of both the recent studies cited, as well as this work. This phenomenon is possibly related to the microscopic effects of exfoliation and emitter generation.

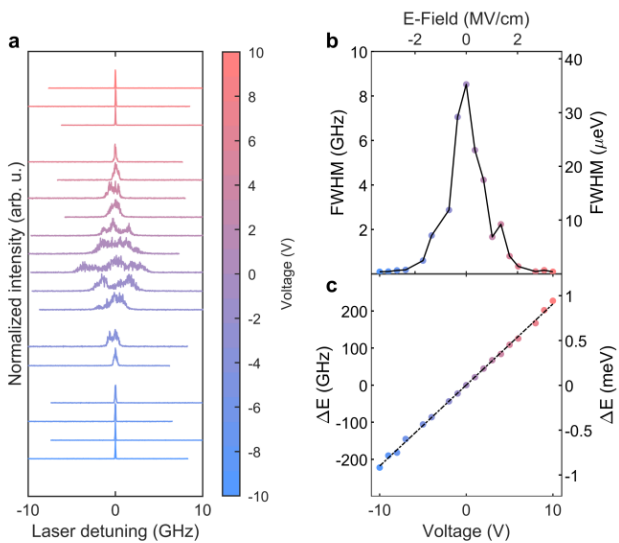

**Figure 3.** (a) PLE spectra of emitter at T=6.5 K for applied voltages ranging from -10V to 10V. (b) Full Width at Half Maximum (FWHM) of the peaks determined with a Voigt profile fit as a function of applied voltage. (c) Measured peak position of the emitter relative to its peak position at 0V as a function of applied voltage.

The effect of temperature on linewidth has been discussed elsewhere [21,23,24] but the linewidth is usually convoluted with inhomogeneous broadening, so the pure effect of temperature on the emitter has been difficult to study independently from the effect of spectral diffusion. Since the electric field can suppress the contribution of the inhomogeneous broadening mechanisms on linewidth, we leveraged this opportunity to study the homogenously broadened linewidth of the h-BN emitter and the effect of temperature on it. We observed a power law increase in the linewidth as temperature increases, with exponent n=3.02 suggesting a $T^3$ dependence of homogeneous linewidth on temperature. This result is in accordance with some earlier studies on h-BN emitters. [23] However, some other studies have shown a linear dependence [21] or a combination of linear and cubic, $AT+BT^3$, [24] dependence of linewidth on temperature. By carefully examining the results of these studies and comparing it with our results, we noticed that the study which used emitters with a strong spectral diffusion effect shows linear dependence [21], and the study which had low temperature linewidth of 1 GHz, exhibits $AT+BT^3$ dependence of linewidth on temperature, with the linear part dominant below 20K [24]. Comparing the results of these studies to our work, in which spectral diffusion is almost completely suppressed, leads us to conclude that spectral diffusion in h-BN is also activated by thermal fluctuations and exhibit a linear variation with temperature, but purely phononic linewidth broadening effects in h-BN result in a $T^3$ dependence of linewidth on temperature. This is in contrast with diamond emitters which show no dependence of spectral diffusion on temperature below 20K [35]

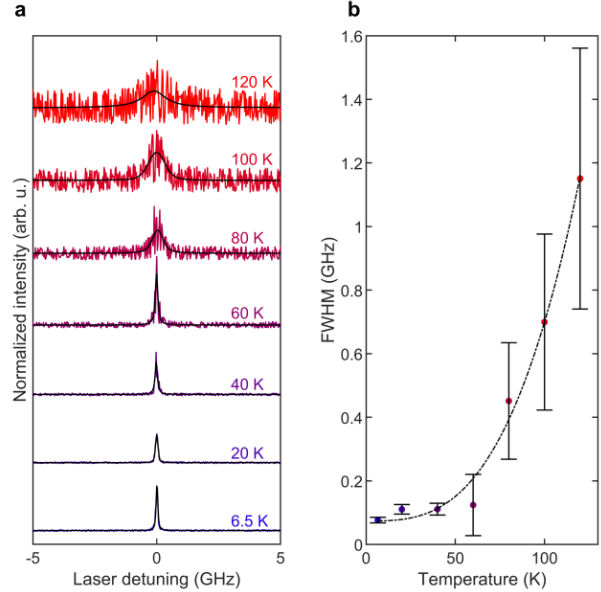

**Figure 4.** (a) PLE traces of the emitter at V=10V for various temperatures. (b) Fitted value of FWHM of the peaks as a function of temperature with a power law fit (FWHM=$AT^n$+C) with a best fit value of n=3.02±0.51, error bars correspond to 95% confidence interval of the fitted values.

A $T^3$ dependence of linewidth on temperature is also observed in diamond nitrogen-vacancy (NV) centers [25]. In the case of NV centers, one possible explanation of the $T^3$ dependence is the fluctuating field created as phonons modulate the distance between the emitter and other defects and impurities, and as a result this type of dependence is mostly observed in samples with high disorder [36]. Our results suggest that this phenomenon is also present in h-BN. The surface of 2D flakes of h-BN is more prone to defects and the emitters in thin layers of h-BN are always close to surface, which can make this the dominant effect in h-BN. This is also in accordance with large spectral diffusion at 0V observed in our study, pointing to the existence of a very defective environment. This argument suggests that in an ideal case in which spectral diffusion is fully suppressed and the defects to modulate local phonons are non-existent, it is possible to have a minimal effect of temperature on linewidth as reported elsewhere [10].

In summary, we have shown that an out-of-plane electrostatic field can suppress spectral diffusion while also Stark tuning the emission energy of hBN color center single photon emitters. Our findings may pave the way for on-chip h-BN quantum communication technologies by removing the barrier to achieving lifetime limited photons, and to potentially generate indistinguishable photons. The extra degree of freedom provided by stark tuning is also desirable to couple the emitters to resonance modes of photonic cavities. Finally, we also studied the dependence of linewidth on temperature in lack of spectral diffusion and found it to be proportional to $T^3$.

**Methods:**

Emitter preparation:

The emitter was prepared by annealing an h-BN crystal (HQ graphene) in between two carbon tablets (MSE supplies Carbon sputtering target >99.99% pure) at 1200C for 10 hours. Then, the h-BN crystal was exfoliated with Scotch tape onto a silicon oxide on silicon substrate. The exfoliated h-BN flake is 13.9 nm thick (see supporting information), and an emitter is identified with a ZPL at 573 nm. This flake containing emitter is used in a device to produce a graphene-hBN-buffer hBN-graphene structure.

Device Fabrication

Graphene, and buffer h-BN were directly exfoliated onto $SiO_2$ (285 nm)/Si substrate and identified via their optical contrast. Substrates were cleaned via ultrasonication in acetone, isopropanol, and deionized water for 20 minutes each and then subject to oxygen plasma at 300mTorr, 70W for 5 minutes. Exfoliation was done at 100°C for higher yield of flakes. All layers were assembled via the polymer-assisted hot pickup technique [37] (With polycarbonate/polydimethylsiloxane stamps). Flakes were picked up between temperatures of 40-70°C and dropped at 180°C. The polycarbonate was washed off in chloroform followed by isopropanol overnight (12 hours). The assembly was done on a substrate of $SiO_2$ (285 nm)/Si with prepatterned electrodes/contacts of 5nm Ti/95nm Au fabricated with electron beam lithography followed by electron beam evaporation of metal and subsequent liftoff in acetone and isopropanol. Finally, the device was wire bonded onto a custom-made printed circuit board. The device was then transferred to Attocube Attodry800 cryo microscope for optical characterization.

Optical characterization:

PL: we use a 532nm CW laser (coherent) as excitation source for PL measurement at 100 µW and used a spectrometer (Princeton Instruments) to measure the PL spectrum. Spectra acquisition time was 60 seconds.

g2 and TRPL: g2 experiment was done at room temperature with 532 nm CW excitation at 100 µW and the light was fiber coupled to a fiber beam splitter (Thorlabs) connected to two avalanche photo diodes (Micro photon devices) and the single photon events were registered by a Picoharp300 electronics (Picoquant) at TTTR mode. The signal was analyzed with a custom MATLAB script

PLE: We used a tunable CW Dye laser (Sirah Mattise), the signal from emitter was passed through a 620 ±10nm bandpass filter and was detected via an avalanche photo diode. The excitation wavelength was measured with a wavemeter (Angstrom). PLE is performed at 10nW power at scan rate of 1 GHz/s.

Acknowledgements:

This work was primarily supported by the 'Photonics at Thermodynamic Limits' Energy Frontier Research Center funded by the U.S. Department of Energy, Office of Science, Office of Basic Energy Sciences under Award Number DE-SC0019140, which supported sample fabrication and optical measurements. Support for system application concepts for hBN emitters has been provide by the Boeing Strategic University program. One of us (H.A.) acknowledges Shahriar Aghaeimeibodi for useful discussions.